\documentclass[sigconf, screen]{acmart}
\newcommand{\eg}[0]{\textit{e.g., }}
\newcommand{\ie}[0]{\textit{i.e., }}
\newcommand{\christine}[1]{{\color{black} #1}}

\newcommand{\ours}{\texttt{MAP~}}
\newcommand{\oursend}{\texttt{MAP}}
\newcommand\jihye[1]{\textcolor{blue!50}{\textbf{Jihye:} #1}}

\usepackage{comment}
\usepackage{stfloats}
\usepackage{main,acmart-taps}
\AtBeginDocument{%
  \providecommand\BibTeX{{%
    \normalfont B\kern-0.5em{\scshape i\kern-0.25em b}\kern-0.8em\TeX}}}

\usepackage[autostyle,german=guillemets]{csquotes}
\makeatletter

\usepackage{hyperref,quoting}
\usepackage{url}
\quotingsetup{vskip=0pt,font={itshape,raggedright},rightmargin=0pt}
\usepackage{tikz}
\usepackage{cuted}
\usepackage{capt-of}
\usepackage{subcaption}
\usepackage{fancyvrb}
\usepackage{gensymb}
\usepackage{float}
\usepackage{stfloats}

\usepackage{graphicx}

\usepackage{enumitem}

\copyrightyear{2025}
\acmYear{2025}
\setcopyright{rightsretained}
\acmConference[CHI EA '25]{Extended Abstracts of the CHI Conference on Human Factors in Computing Systems}{April 26-May 1, 2025}{Yokohama, Japan}
\acmBooktitle{Extended Abstracts of the CHI Conference on Human Factors in Computing Systems (CHI EA '25), April 26-May 1, 2025, Yokohama, Japan}\acmDOI{10.1145/3706599.3719853}
\acmISBN{979-8-4007-1395-8/2025/04}

\begin{document}

\title[\oursend: Multi-user Personalization with Collaborative LLM-powered Agents]{\oursend\,\raisebox{-0.2em}{\includegraphics[height=1.1em]{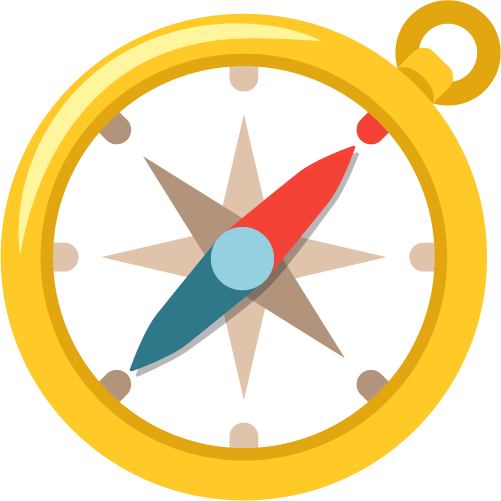}}\,: Multi-user Personalization with \\ Collaborative LLM-powered Agents}


\author{Christine P. Lee}
\orcid{0000-0003-0991-8072}
\authornote{Both authors contributed equally to this research.}
\affiliation{%
  \institution{Department of Computer Sciences University of Wisconsin--Madison}
  \country{Madison, Wisconsin, USA}
}
\email{cplee5@cs.wisc.edu}

\author{Jihye Choi}
\orcid{0009-0000-9719-3758}
\authornotemark[1]
\affiliation{%
  \institution{Department of Computer Sciences University of Wisconsin--Madison}
  \country{Madison, Wisconsin, USA}
}
\email{jihye@cs.wisc.edu}

\author{Bilge Mutlu}
\orcid{0000-0002-9456-1495}
\affiliation{%
  \institution{Department of Computer Sciences University of Wisconsin--Madison}
  \country{Madison, Wisconsin, USA}
}
\email{bilge@cs.wisc.edu}
\renewcommand{\shortauthors}{}

\begin{abstract}
The widespread adoption of Large Language Models (LLMs) and LLM-powered agents in multi-user settings underscores the need for reliable, usable methods to accommodate diverse preferences and resolve conflicting directives. 
Drawing on conflict resolution theory, we introduce a user-centered workflow for multi-user personalization comprising three stages: \textit{Reflection}, \textit{Analysis}, and \textit{Feedback}. We then present \oursend—a \textbf{M}ulti-\textbf{A}gent system for multi-user \textbf{P}ersonalization—to operationalize this workflow.
By delegating subtasks to specialized agents, \oursend\ (1) retrieves and reflects on relevant user information, while enhancing reliability through agent-to-agent interactions, (2) provides detailed analysis for improved transparency and usability, and (3) integrates user feedback to iteratively refine results. Our user study findings ($n=12$) highlight \oursend's effectiveness and usability for conflict resolution while emphasizing the importance of user involvement in resolution verification and failure management. 
This work highlights the potential of multi-agent systems to implement user-centered, multi-user personalization workflows and concludes by offering insights for personalization in multi-user contexts.

\end{abstract}


\maketitle

\section{Introduction}
\label{sec:intro}

As artificial intelligence (AI) continues to advance, its deployment in software, smart products, and robots is expanding into increasingly complex real-world scenarios \cite{sullivan2024making, kuo2023understanding, kim2024understanding}. 
These AI systems are becoming more autonomous in perceiving and interacting with their surroundings, exhibiting qualities such as reactivity, social engagement, and the ability to learn and adapt \cite{cila2022designing, xi2023rise}. 
In practice, these AI systems are often introduced into \textit{multi-user environments}, such as smart home devices for families, automated scheduling for teams, and assistive robots in shared living spaces \cite{lee2024rex, seeber2020machines, mutlu2008robots}. 
These contexts require AI systems to accommodate multiple users by integrating diverse individual preferences and instructions into a cohesive decision-making process. This highlights the need for AI to effectively manage complexity and adapt for seamless integration into such environments.

Imagine an AI-powered household robot receiving conflicting requests from different family members~\cite{lee2024rex}---one user asks to keep the kitchen counter untouched due to recent caulking, while another unknowingly requests groceries to be placed on the same counter. In such situations, an ideal adaptive response from the robot would involve reconciling these conflicting instructions by recalling previously stated preferences, aligning its actions with the collective needs of all users. 
Additionally, the robot could generate explanations for its altered actions and accept updated instructions regarding the kitchen counter or groceries. Such adaptability can also enhance the long-term usability of AI systems by storing and retrieving user preferences when needed, reducing the need for users to repeatedly express their preferences or reestablish context.

For AI systems to effectively adapt and personalize in multi-user environments, they require clear frameworks that guide their personalization efforts. These frameworks should enable AI systems to engage in ongoing learning, accommodate diverse user requests, and adjust their operations to meet evolving needs and preferences.
Moreover, they must consider the unique complexities introduced by multi-user dynamics. Addressing overlapping or conflicting preferences requires decision-making, prioritization, and, at times, the abandonment of certain requests. Successfully managing these conflicts and achieving resolutions that ensure a satisfactory and accepted user experience is a crucial challenge as AI systems become more integrated into multi-user settings.

In this work, we aim to develop guidelines and demonstrate our approach to providing personalized experiences with AI systems in multi-user environments. Our approach leverages recent advancements in large language models (LLMs), which have proven to be powerful tools for developing adaptive personalization solutions and aligning with human preferences through methods such as reward function modeling~\cite{kwon2023reward, yu2023language} and fine-tuning~\cite{bai2022rlhf, rafailov2024dpo}.
\christine{Specifically, we utilize an orchestration of multiple LLM agents to collaborate on complex conflict resolution and preference adaptation tasks \cite{guo2024large, gao2024large}.} 
Our research is guided by the central question: \textit{``How should LLM-powered AI systems be designed and built to provide adaptive personalization for multiple users?''} To address this question, we introduce a user-centered workflow tailored for multi-user personalization, structured around the unique challenges of multi-user environments—particularly conflict resolution. Drawing from conflict resolution literature, we propose a three-stage framework: reflection, analysis, and feedback.
We then present \ours (\textbf{M}ulti-\textbf{A}gent for Multi-User \textbf{P}ersonalization), an implementation of our workflow using a collaborative orchestration of LLM-powered agents. We then assess the effectiveness of our designed workflow and \ours through a user study to understand user experiences and design requirements for AI systems deployed in multi-user settings.\footnote{We provide an extended discussion on related literature regarding personalization with LLMs and LLM-based agents in Appendix \ref{app:related}.}


\section{Interaction Design of \oursend}
This section outlines the interaction design of \oursend, focusing on the challenges and requirements of multi-user personalization. We first differentiate single-user and multi-user personalization to highlight the complexities of accommodating multiple users (Section \ref{sec:single}). We then introduce real-world scenarios illustrating these challenges (Section \ref{sec:scenarios}). Finally, we present a structured workflow---\textit{Reflection}, \textit{Analysis}, and \textit{Feedback}---to guide \oursend's approach to managing personalization and conflict resolution (Section \ref{sec:fw2}).
\label{sec:prelim}
\subsection{Single-user vs Multi-user Personalization}\label{sec:single}
Here we define our target task, highlighting the key differences between single-user and multi-user personalization.
The tasks and interaction requirements for human-AI interactions differ significantly depending on whether the personalization task involves a single user or multiple users.
In \textit{single-user} interactions, the system focuses on adapting to the preferences and rules of an individual. 
The primary goal is to efficiently manage and personalize tasks to meet the unique needs of the individual without encountering external conflicts.
In contrast, \textit{multi-user} interactions introduce a more complex dynamic. The system must accommodate diverse and often conflicting preferences from multiple users. Requirements extend beyond basic personalization to include the identification and resolution of conflicts between users' preferences and schedules. This creates a more complicated circumstances, where balancing competing needs and resolving conflicts become central tasks, leading to more sophisticated requirements for the AI system.

\subsection{Everyday Scenarios} \label{sec:scenarios} 
Throughout this paper, we focus on scenarios designed to simulate the multi-user personalization task, representing situations that commonly arise in daily life among multiple users. Each scenario involves interactions between an AI system and three hypothetical users, each with a weekly schedule and a set of preferences dictating how specific tasks should be handled. The objective in each scenario is for the AI system or users to generate daily plans that effectively accommodate the diverse schedules and preferences of all users. 
Although these scenarios are not exhaustive or highly sophisticated, they aim to provide a meaningful and representative foundation for examining complex multi-user tasks. Extended scenario descriptions can be found in Appendix~\ref{app:scenarios-extended}.

\begin{enumerate}
\item \textit{Scenario 1: Workplace Scheduling.} An AI system schedules meeting rooms for a consulting firm, balancing employee preferences (\eg ambiance, lighting, seating) and resolving conflicts by prioritizing client consultations, team meetings, brainstorming, and other activities in order.

\item \textit{Scenario 2: Assistive Care Robot.} \label{sec:scenarios2} An AI-powered robot in an assistive care facility manages residents' preferences and schedules, such as providing interactive wake-up calls, mobility assistance, and custom deliveries. Conflicts are resolved by prioritizing requests alphabetically by residents' names.

\item \textit{Scenario 3: Smart-home Temperature Control.} A smart-home AI regulates communal area temperatures based on housemates' schedules and preferences, such as warm settings for study or cooler ones for cost-saving. Conflicts are resolved through discussions among housemates.
\end{enumerate}



\label{sec:workflow}
\subsection{Multi-user Personalization Workflow: Reflection, Analysis, and Feedback}
\label{sec:fw2}

Managing conflicts among users in multi-user personalization systems presents unique challenges, such as understanding the dynamic nature of user preferences, \christine{facilitating coordination} in resolving conflicts, and providing transparency in decision-making processes. Addressing these complexities necessitates a principled approach that enables the systems to consistently provide conflict resolution strategies under evolving conditions. 
To this end, we draw on principles from the human conflict resolution theory~\cite{thyagaraju2008conflict, rothman1997action} to develop technical solutions to effectively mediate and resolve conflicts.

Specifically, \citet{ross2001action} introduces an \textit{Action Evaluation} methodology that integrates goal setting, monitoring, and evaluation directly into the conflict resolution process. This methodology is structured around three iterative stages: reflection, analysis, and feedback. Adopting this structure, we propose a three-stage workflow for AI systems managing multi-user personalization:


\begin{enumerate}
    \item \textit{Reflection}: The system collects and aggregates personalization elements from various users, including preferences, interaction history, and schedules. This stage mirrors the participatory nature of Action Evaluation by involving all stakeholders in defining their needs and goals.

    \item \textit{Analysis}: The system synthesizes the collected data to construct a comprehensive plan that adheres to multiple users' needs. When conflicts arise, the system generates resolutions that account for individual preferences and shared objectives. The final plan is presented to users, detailing the decision-making process, the rules considered, conflicts detected, and the proposed resolutions. This stage parallels the intervention planning in Action Evaluation, which incorporates mechanisms to address disagreements to achieve goals defined in the reflection stage.

    \item \textit{Feedback}: The system gathers user feedback on the generated plan. This feedback may include adjustments to existing settings, updates to user-defined preferences, or suggestions to refine conflict resolution strategies. The system incorporates this feedback into its planning, for iterative improvement and alignment with user needs over time. This continuous refinement embodies the feedback stage of Action Evaluation, which revisits and refines the goals to ensure continued progress and alignment with participants' evolving needs. 
\end{enumerate}

By incorporating these stages, the proposed workflow ensures that AI systems managing multi-user personalization can mediate conflicts, adapt to dynamic user needs, and \christine{uphold transparency} and satisfaction across diverse stakeholders. Building on Action Evaluation's principles of iterative goal setting and evaluation, our approach emphasizes the importance of transparency, adaptability, and user engagement in AI-mediated conflict resolution.

\section{\oursend: Multi-Agent LLMs for Multi-User Personalization}

\label{sec:system}
\begin{figure*}[!t]
  \includegraphics[width=\textwidth]{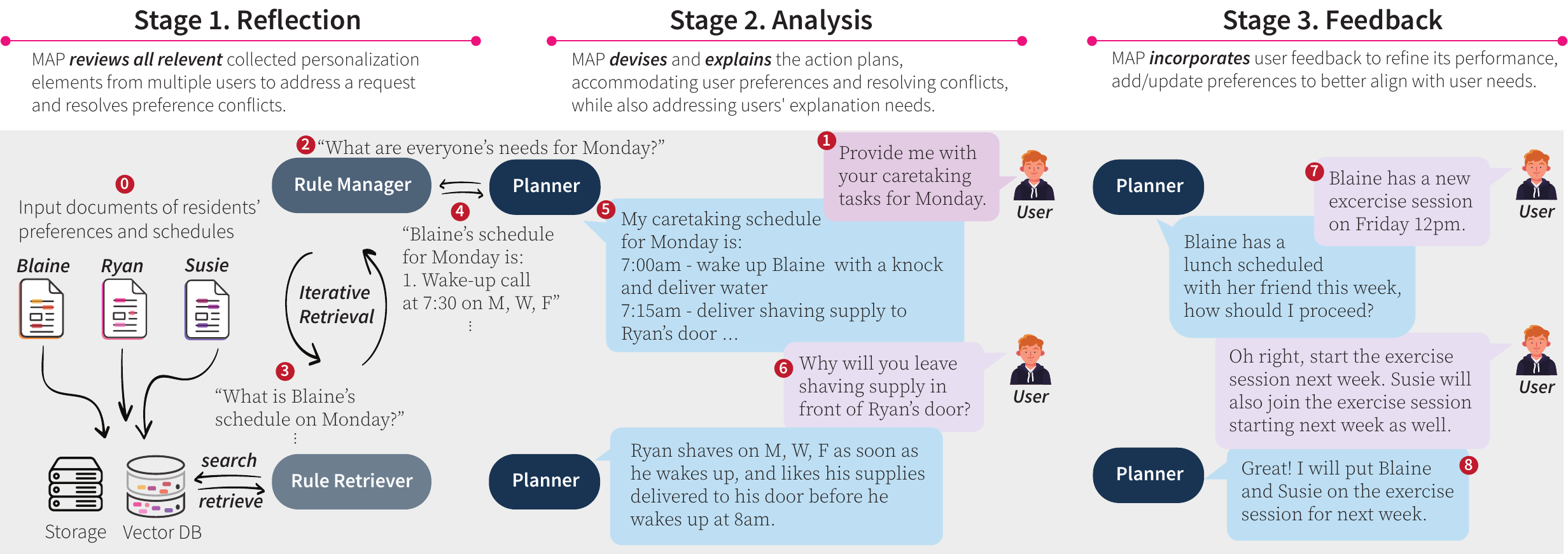}
  \caption{
  \textit{Our multi-agent system to support the proposed multi-user personalization workflow ---}
  The system orchestrates specialized agents through three stages of our user-centered workflow—Reflection, Analysis, and Feedback—to retrieve user data, reason about personalization tasks, resolve conflicts, and incorporate user feedback.
  }

  \label{fig:rag}
\end{figure*}



Recent advances in LLMs have led to the emergence of agents capable of handling complex tasks \cite{wu2023autogen, du2023improving, madaan2024selfcritic, choi2024malade, hong2024metagpt}. 
In contrast to a static, standalone LLM, an agent is a dynamic, intelligent entity that leverages an LLM’s capabilities without task-specific training, by incorporating function-calling/tools or external knowledge bases. By coordinating multiple agents, each tasked with a specific subtask, these systems work collaboratively to address more complex, nuanced end-tasks such as multi-user personalization. 
Here we propose~\oursend, a multi-agent LLM-based system that supports user-centered workflows introduced in Section~\ref{sec:workflow}: Reflection, Analysis, and Feedback. Refer to Appendix~\ref{app:system-details} for further details.\footnote{The code repository for our work is available at~\url{https://github.com/jihyechoi77/MAP}.}

At the core of~\ours is an agent called the \textit{Planner}, which serves as the highest-level interface between the user and the system.
When given a user query such as ``Provide me with your caretaking tasks for Monday'' (step \redcircle{1} in Figure~\ref{fig:rag}), the system must (1) collect all relevant information to answer, and (2) reason about this information and generate the personalized response to the user.
Rather than assigning both (1) and (2) exclusively to the \textit{Planner}, we offload (1) to specialized supporting agents (\ie \textit{Rule Retriever} and \textit{Rule Manager}). This division frees the \textit{Planner} to focus on (2)~\footnote{Such a (1) vs. (2) distinction draws on recent insights from the LLM uncertainty literature \cite{houdecomposing, tobelieveeornot, ling2024uncertainty, yin-etal-2024-reasoning}, which identifies two key factors determining the reliability of LLM outputs: (1) the quality of demonstrations in the prompt, and (2) the LLM’s internal reasoning capabilities. Note that (1) corresponds to the Reflection phase, while (2) corresponds to the Analysis and Feedback phases in our multi-user personalization workflow.}.

\paragraph{Stage 0: Data Ingestion.} Initially, users are asked to furnish their preferences in documents (depicted as step \redcircle{0}).
These documents contain information pertinent to each user, encompassing individual requests, schedules, or rules. 
Subsequently, a pre-processing, or \textit{ingestion phase} ensues, wherein documents are not just stored in a storage but also sharded into reasonable-size chunks and ingested into a suitable type of document store (\eg vector database).
This document store serves as a form of external storage where all users' personal information and preferences are stored, which is desired to be incorporated into the interactions of human users with the~\ours system.

\paragraph{Stage 1: Reflection.}
To instantiate the reflection phase of the personalization workflow, the Planner agent is scheduled to collect all relevant user information before providing any personalized plans to the user prompts (step \redcircle{2}). 
Rule Retriever and Rule Manager agents effectively support this process. 
Rule Retriever, an LLM-powered agent with Retrieval Augmented Generation (RAG), retrieves a requested piece of information by directly accessing the document store from the data ingestion phase; it retrieves the top-$k$ most \textit{relevant} document-chunks to a given query.
To further enhance retrieval quality, Rule Retriever is managed by another LLM-powered agent, called Rule Manager. 
Rule Manager breaks the Planner's query into more granular, specifically executable subqueries (\eg ``What is Blaine's schedule?'') and iteratively sends each of them to the Rule Retriever (\redcircle{3}) until all information is gathered (\redcircle{4}).

\paragraph{Stage 2: Analysis.}
After receiving the compiled information from the Rule Manager in Stage 1, the Planner generates personalized answers to the original query by the user (step \redcircle{5}).
In doing so, it resolves any conflicting preferences among multiple users, and articulates the reasoning behind its responses; providing a summary of relevant rules referred to, upon request (\redcircle{6}). Prompts used for the Planner can be found in Appendix~\ref{app:PlannerPrompt}.

\paragraph{Stage 3: Feedback.}
The user offers feedback on the Planner’s response (step \redcircle{7}). The Planner maintains a conversation history, leveraging a large context window to keep track of previous queries and feedback. Future iterations could further improve personalization by ingesting new user inputs or additional context back into the document store (\redcircle{8}).


\begin{figure*}[!t]
  \includegraphics[width=\textwidth]{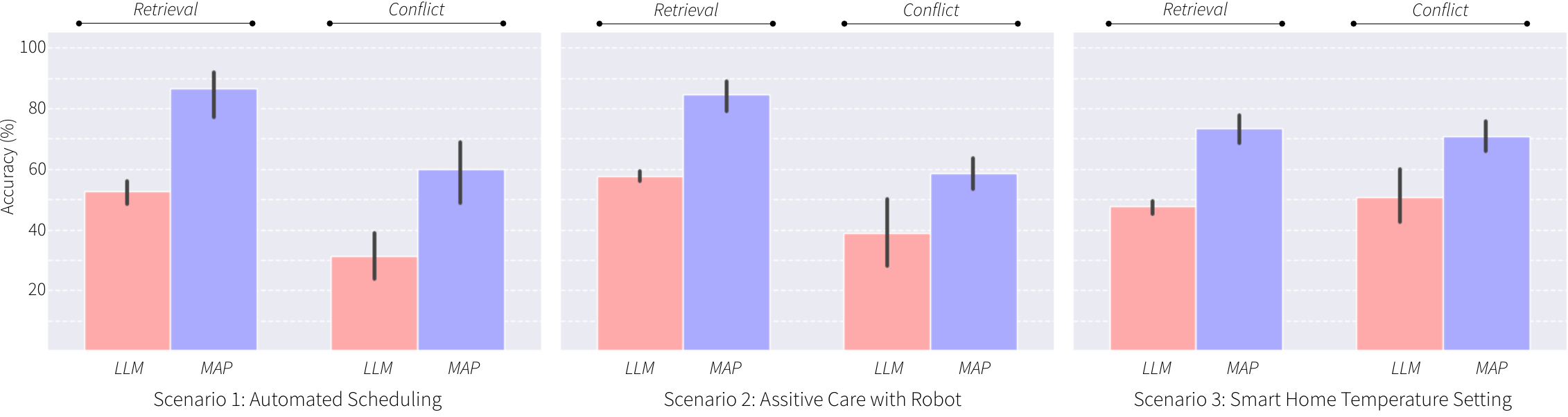}
   \vspace{-12pt}
   \caption{\textit{Monolithic LLM-based approach (``LLM'') vs. our multi-agent approach (``MAP'') across the three multi-user personalization scenarios---} The evaluation measures each system's success (in \%) of 1) completely retrieving all 60 user preferences and schedules to generate personalized action plans (``Retrieval''), and 2) identifying and resolving all 12 present conflicts among the users (``Conflict'').
}
  \label{fig:eval}
\end{figure*}

\subsection{Limitations of a Monolithic LLM Approach for Multi-User Personalization}
Recent LLMs~\cite{gpt-4, geminiteam2023gemini, claude} are well known for their long-enough context window, advanced reasoning capabilities, and handling nuanced language—qualities that seem well-suited for complex personalization tasks. 
One might therefore wonder if a basic chat agent, implemented as a monolithic LLM with a conversation history, could sufficiently support the proposed user-centric workflow in Section~\ref{sec:workflow}.
However, we emphasize the advantages of a modular design, in which the overarching task is decomposed into subtasks that are distributed among multiple agents. Such a composition-based approach not only enhances overall system reliability (as we demonstrate in the quantitative evaluation below) but also provides flexibility to adapt to evolving changes in the personalization workflow. As the complexity of personalization grows—with more users, more intricate rules, or additional user feedback—an architecture composed of specialized agents can more readily incorporate new features (\eg failure-handling functionality as discussed in Section~\ref{sec:study-findings}) without requiring extensive rewrites to the entire system.


Furthermore, in Figure~\ref{fig:eval}, we compare the effectiveness of a monolithic LLM (LLM, red bars) and our multi-agent framework (\oursend, blue bars) across three personalization scenarios described in Section~\ref{sec:scenarios}. For the monolithic approach, we instantiate a single GPT-4-based chat agent. Our evaluation focuses on two main criteria: (1) \textit{Retrieval}, reflecting how comprehensively the system accounts for user rules; and (2) \textit{Conflict Resolution}, measuring how well it identifies and resolves conflicts among multiple users-a core challenge in multi-user personalization. 
In our evaluation, we assessed the accuracy of both the single-agent LLM and \oursend. Each scenario involved 60 rules governing the preferences and schedules of three users, leading to 12 conflicts due to overlapping preferences and schedules.
Each model was tested three times across three scenarios. Since each test generated five daily action plans for weekdays, this resulted in 15 instances per scenario and 45 instances per condition, totaling 90 instances for validation.
We then compared each generated output against a reference solution formulated by the researcher who designed the scenarios. To evaluate accuracy, we measured errors where the models failed to correctly align with the intended schedules or preferences.
The results in Figure~\ref{fig:eval} indicate that \ours significantly outperforms the monolithic LLM across all scenarios.
This performance gap arises in part because the monolithic LLM’s ability to resolve conflicts is hampered by its limited retrieval performance.
For instance, with a limited retrieval rate ($\sim50\%$) of user preferences and schedules, it often fails to generate responses that fully account for all user rules, leading to inadequate conflict resolution ($\sim30\%$ at the lowest). 
Typical errors include overlooking scheduling overlaps (\eg multiple users requesting the same meeting room at 12:00 pm on Friday) and ignoring conflicting requests (\eg one user’s temperature ceiling of 70\degree C versus another’s preference for 80\degree C). By contrast, the multi-agent architecture ensures more comprehensive retrieval and thus more robust conflict resolution.



\section{User Study of \ours}
\label{sec:study}

\subsection{Study Design}

We conducted a user study to evaluate the effectiveness of the multi-user personalization workflow and \oursend. The study used a within-subjects design with scenarios outlined in Section \ref{sec:scenarios} as the within-subjects variable. Participants were introduced to \oursend's workflow stages (reflection, analysis, and feedback) and given examples to familiarize themselves with its interaction process. Each participant was assigned two of three scenarios in a counterbalanced order and tasked with creating weekly plans considering varied preferences and scheduling requirements.

Participants interacted with \ours through its web interface, navigating the workflow stages and evaluating the system's proposed plans and conflict resolution strategies. Qualitative data was gathered via semi-structured interviews conducted in person, with sessions recorded on Zoom \cite{zoom2023}. Each session lasted approximately one hour, and participants were compensated \$15 per hour. Session recordings were transcribed using Otter.ai \cite{OtterAI2023} and manually reviewed. Qualitative analysis followed thematic analysis \cite{clarke2014thematic, McDonald19}, with open coding performed by the first author to identify key concepts, which were then clustered into themes. The research team iteratively refined the themes through discussion and consensus.

\subsubsection{Participant Demographics}
12 participants (P1–P12) were recruited through university mailing lists for our user study. Eligibility criteria included being in the United States, fluent in English, and at least 18 years old. Participants ranged in age from 20 to 56 years ($M = 30.6$, $SD = 10.8$), with 58.4\% identifying as female and 41.6\% as male. The racial composition was 83.4\% White, 8.3\% Asian, and 8.3\% preferring not to respond. Findings are reported using the notation P\textit{i}, where \textit{i} represents the participant ID.

\subsection{Findings}
\label{sec:study-findings}
Our analysis provided insights into participants' perceptions of the proposed workflow and their interaction experience with \oursend. Below, we present the key themes identified in our findings.


\subsubsection{Conflict Resolution as an AI-Suitable Task in Multi-user Settings}

Ten participants (P1–P5, P7, P8, P10–P12) described conflict detection and resolution as a suitable task for AI, as it is an unfavorable and undesired responsibility for humans. They explained that conflicts in multi-user settings are inevitable, as differences in preferences and overlapping schedules frequently arise, and users are often unaware of each other's needs and potential conflicts.

Participants further elaborated that resolving such conflicts among multiple people is mentally and emotionally draining. The cognitive load of tracking various preferences and schedules, coupled with the challenge of predicting potential conflicts, makes the task complicated and time-consuming. As P4 expressed: \textit{``I can't possibly keep track of what everyone in the house prefers, nor predict every conflict when planning things. It's overwhelming. Having the system not only identify these conflicts but also resolve them for me is a huge relief---it saves so much time and prevents unnecessary arguments or assumptions.''} 
Consequently, participants found that analyzing such information and generating mitigating plans is an ideal task for the AI system. Moreover, they envisioned that the role of the system in resolving conflicts could be adaptively used even in high-stakes tasks, where it could detect conflicts and suggest solutions, while still allowing humans to confirm or adjust the final resolution.

\subsubsection{Reflection Stage Enhancing Usability}

Participants highlighted that the system's ability to save user preferences and automatically retrieve information to generate plans significantly improved usability. This improvement in usability stemmed from two key aspects: (1) the automatic creation of plans for complex specifications and (2) the ease of not having to prompt manually.

Eleven participants (P1-P10, P12) reported that the system's management and generation of schedules and preferences reduced their effort and time, easing the mental load of integrating diverse inputs. P3 explained: \textit{``I can't hold all the preferences in my head, so it was really helpful that the system organized [preferences]. All I had to do was confirm and think of new fun rules to add. When I did it manually, I struggled just to ensure basic rules were met. The system really took the load off me.''} While manual management was feasible, participants noted that accommodating diverse preferences was particularly challenging without automation.

Four participants (P3, P7, P11, P2) further emphasized that the system's ability to retrieve information without requiring repeated user input streamlined the interaction experience. This feature allowed users to build on existing information rather than repeatedly entering extensive preferences and schedules. Participants envisioned such functionality enhancing usability in real-world settings, enabling users to expand upon prior lists with minimal effort.

\subsubsection{Analysis and Feedback Stage Supporting Transparency and Adaptability}

Eight participants (P2, P3, P4, P6–P9, P12) emphasized the importance of the analysis and feedback stages, particularly working together to achieve effective personalization. They highlighted that the analysis stage was important for system transparency and output verification, while the feedback stage enabled adaptability to address the needs identified during analysis.

The analysis stage provided accessible and understandable information through \oursend's interface and natural language capabilities, helping participants clarify the system's decision-making logic and data sources (\eg other users' preferences and schedules) based on their needs. Participants also valued the analysis stage for verifying the system's outputs and logic. Four participants (P3, P6, P7, P12) specifically used these explanations to better understand the decision-making process by comparing outputs against the provided list of rules and resolving uncertainties. They emphasized that such explanations were key to building trust with the system. For instance, P6 remarked: \textit{``As I asked more questions and received additional details, it clarified my doubts and blind spots. I felt more confident, thinking, `Okay, I can trust this system and try more complex tasks, but not too extreme ones, since I can always ask questions and provide feedback if needed.' ''} By fostering transparency and verification of planning and resolution outcomes, the analysis stage played a crucial role in enhancing user acceptance.

Building on the insights gained from the analysis stage, participants applied this information to adapt \oursend' behavior to their individual or group needs and preferences. Seven participants (P1, P2, P3, P8–P11) noted that the feedback stage was key to ensuring adaptability, allowing the system to respond to evolving needs and group norms. They highlighted the dynamic nature of preferences and schedules, with P9 explaining: \textit{``The initial preferences are never set in stone. Things come up, and if a system designed to help me can't change with me, it's not very helpful.''} Some participants (P2, P3, P8–P10) further emphasized the role of feedback in aligning the system with group-specific norms and priorities. As P10 observed: \textit{``These rules reflect what we, as a group, want and value, so the system needs to follow and blend into our preferences,''} highlighting the feedback mechanism's importance in addressing both individual and collective needs.

\subsubsection{Support Needed for Retrieval and Conflict Resolution Failures}

During system engagement, participants (P1, P3, P4, P6, P9, P11, P12) identified issues where the system failed to retrieve relevant rules, misdetected conflicts, or generated inconsistent resolutions. For example, one participant noted, \textit{P1: ``Sometimes it followed my preferences at the expense of my roommate's, and other times it did the opposite. This unpredictability makes it hard to work with or plan around.''} While participants found the automated planning and conflict resolution features useful, these inconsistencies raised concerns about overreliance on the system and skepticism in the reliabilty of the resolution capabilities. 


To address these issues, participants expressed a need for robust explanations and verification mechanisms. They wanted clarity on which rules were considered, which were missed, and how conflicts were resolved. One participant highlighted, \textit{P9: ``Given that it's complicated and likely to make choices [in what rules to choose], I want to just check what choices were made, to make sure its the best choice for all of us.''} Another suggested the system provide comparisons between processed rules and the full rule list, flagging omissions. \textit{P4: ``It would be helpful if the system could flag where it's unsure about certain preferences, so I know to pay closer attention to those areas.''} These features were seen as critical for navigating system shortcomings and enabling better-informed decision-making.

\section{Discussion and Future Work}
\label{sec:conclusion}
In this work, we focus on enabling AI systems to provide personalization in multi-user settings, where unique and complex dynamics often lead to conflicting preferences and needs. To address these challenges, we propose a workflow inspired by conflict resolution theory, which guides AI systems through three critical stages: reflection, analysis, and feedback. 
Our findings demonstrate that this three-stage workflow effectively facilitates transparent and adaptive personalization while efficiently managing the complexity of balancing multiple users' needs. Each stage plays a crucial role in cooperatively navigating and addressing multi-user demands, showing promise for extension to more complex scenarios.

Our work highlights the importance of embedding a user-centric workflow into the design of multi-agent systems. By starting with the workflow requirements (\ie how humans resolve conflicts in the real-world) and then constructing agents around those requirements, we show how AI systems can better serve users---particularly in critical domains like multi-user personalization. \christine{However, as our system revealed technical limitations, such as retrieval failures, the workflow can be adapted to incorporate user control where needed while leveraging LLM capabilities effectively. For instance, to mitigate retrieval failures, the system can generate multiple retrieval attempts and plans, then incorporate user input to rank the options or provide feedback for refinement.}

Our study also revealed the need for greater user involvement and verification tools within the workflow, as well as mechanisms to address conflict resolution failures. 
Future work could explore enhancing the MAP framework with additional agents that further engage users, for example through critic or validator roles~\cite{choi2024malade, madaan2024selfcritic}. Such agents could monitor and correct the Planner’s outputs, ensure responses adhere to specific user constraints, or invoke user feedback and external function calls to adjust rules, preferences, or weighting schemes~\cite{han-etal-2024-towards}.
\christine{In addition, programming language methods such as formal verification and repair can be integrated into LLMs (\eg \cite{lee2025veriplan}). Formal verification techniques, such as model checking, can impose deterministic boundaries to counteract the probabilistic nature of LLMs. These boundaries, representing user-defined preferences and rules, can ensure that LLM outputs strictly adhere to specified constraints. Such methods could address the inconsistent conflict resolutions observed in \ours by providing structured guidelines for coordination. Future work can explore applying formal methods, like program synthesis, verification, and repair, to enhance LLM reliability and alignment with user needs.}

Another avenue for investigation is extending these ideas to larger-scale personalization scenarios---incorporating more users, more complex preference structures, and additional types of task conflicts. By broadening the scope, researchers can explore both the scalability of our proposed agentic approach and the reliability of its conflict resolution processes.



\begin{acks}
We thank the reviewers for their helpful comments. We also thank the participants in our study. This work was supported by the National Science Foundation awards 1925043 and 2152163. 
Jihye Choi is partially supported by DARPA under agreement number 885000, NSF CCF-FMiTF-1836978 and ONR N00014-21-1-2492. 
\end{acks}

\balance
\bibliographystyle{ACM-Reference-Format}
\bibliography{bibliography}


\begin{thebibliography}{53}


\ifx \showCODEN    \undefined \def \showCODEN     #1{\unskip}     \fi
\ifx \showDOI      \undefined \def \showDOI       #1{#1}\fi
\ifx \showISBNx    \undefined \def \showISBNx     #1{\unskip}     \fi
\ifx \showISBNxiii \undefined \def \showISBNxiii  #1{\unskip}     \fi
\ifx \showISSN     \undefined \def \showISSN      #1{\unskip}     \fi
\ifx \showLCCN     \undefined \def \showLCCN      #1{\unskip}     \fi
\ifx \shownote     \undefined \def \shownote      #1{#1}          \fi
\ifx \showarticletitle \undefined \def \showarticletitle #1{#1}   \fi
\ifx \showURL      \undefined \def \showURL       {\relax}        \fi
\providecommand\bibfield[2]{#2}
\providecommand\bibinfo[2]{#2}
\providecommand\natexlab[1]{#1}
\providecommand\showeprint[2][]{arXiv:#2}

\bibitem[Abbasi-Yadkori et~al\mbox{.}(2024)]%
        {tobelieveeornot}
\bibfield{author}{\bibinfo{person}{Yasin Abbasi-Yadkori}, \bibinfo{person}{Ilja
  Kuzborskij}, \bibinfo{person}{Andr{\'a}s Gy{\"o}rgy}, {and}
  \bibinfo{person}{Csaba Szepesvari}.} \bibinfo{year}{2024}\natexlab{}.
\newblock \showarticletitle{To Believe or Not to Believe Your {LLM}:
  IterativePrompting for Estimating Epistemic Uncertainty}. In
  \bibinfo{booktitle}{\emph{The Thirty-eighth Annual Conference on Neural
  Information Processing Systems}}.
\newblock
\urldef\tempurl%
\url{https://openreview.net/forum?id=k6iyUfwdI9}
\showURL{%
\tempurl}


\bibitem[Achiam et~al\mbox{.}(2023)]%
        {gpt-4}
\bibfield{author}{\bibinfo{person}{Josh Achiam}, \bibinfo{person}{Steven
  Adler}, \bibinfo{person}{Sandhini Agarwal}, \bibinfo{person}{Lama Ahmad},
  \bibinfo{person}{Ilge Akkaya}, \bibinfo{person}{Florencia~Leoni Aleman},
  \bibinfo{person}{Diogo Almeida}, \bibinfo{person}{Janko Altenschmidt},
  \bibinfo{person}{Sam Altman}, \bibinfo{person}{Shyamal Anadkat},
  {et~al\mbox{.}}} \bibinfo{year}{2023}\natexlab{}.
\newblock \showarticletitle{Gpt-4 technical report}.
\newblock \bibinfo{journal}{\emph{arXiv preprint arXiv:2303.08774}}
  (\bibinfo{year}{2023}).
\newblock


\bibitem[Anil et~al\mbox{.}(2023)]%
        {geminiteam2023gemini}
\bibfield{author}{\bibinfo{person}{Rohan Anil}, \bibinfo{person}{Sebastian
  Borgeaud}, \bibinfo{person}{Yonghui Wu}, \bibinfo{person}{Jean-Baptiste
  Alayrac}, \bibinfo{person}{Jiahui Yu}, \bibinfo{person}{Radu Soricut},
  \bibinfo{person}{Johan Schalkwyk}, \bibinfo{person}{Andrew~M. Dai},
  \bibinfo{person}{Anja Hauth}, \bibinfo{person}{Katie Millican},
  \bibinfo{person}{David Silver}, \bibinfo{person}{Slav Petrov},
  \bibinfo{person}{Melvin Johnson}, \bibinfo{person}{Ioannis Antonoglou},
  \bibinfo{person}{Julian Schrittwieser}, \bibinfo{person}{Amelia Glaese},
  \bibinfo{person}{Jilin Chen}, \bibinfo{person}{Emily Pitler},
  \bibinfo{person}{Timothy Lillicrap}, \bibinfo{person}{Angeliki Lazaridou},
  \bibinfo{person}{Orhan Firat}, {et~al\mbox{.}}}
  \bibinfo{year}{2023}\natexlab{}.
\newblock \bibinfo{title}{Gemini: A Family of Highly Capable Multimodal
  Models}.
\newblock
\newblock
\showeprint[arxiv]{2312.11805}~[cs.CL]


\bibitem[Anthropic(2024)]%
        {claude}
\bibfield{author}{\bibinfo{person}{Anthropic}.}
  \bibinfo{year}{2024}\natexlab{}.
\newblock \showarticletitle{The Claude 3 Model Family: Opus, Sonnet, Haiku}.
\newblock  (\bibinfo{year}{2024}).
\newblock
\urldef\tempurl%
\url{https://www-cdn.anthropic.com/de8ba9b01c9ab7cbabf5c33b80b7bbc618857627/Model_Card_Claude_3.pdf}
\showURL{%
\tempurl}


\bibitem[Bai et~al\mbox{.}(2022)]%
        {bai2022rlhf}
\bibfield{author}{\bibinfo{person}{Yuntao Bai}, \bibinfo{person}{Andy Jones},
  \bibinfo{person}{Kamal Ndousse}, \bibinfo{person}{Amanda Askell},
  \bibinfo{person}{Anna Chen}, \bibinfo{person}{Nova DasSarma},
  \bibinfo{person}{Dawn Drain}, \bibinfo{person}{Stanislav Fort},
  \bibinfo{person}{Deep Ganguli}, \bibinfo{person}{Tom Henighan},
  {et~al\mbox{.}}} \bibinfo{year}{2022}\natexlab{}.
\newblock \showarticletitle{Training a helpful and harmless assistant with
  reinforcement learning from human feedback}.
\newblock \bibinfo{journal}{\emph{arXiv preprint arXiv:2204.05862}}
  (\bibinfo{year}{2022}).
\newblock


\bibitem[Brown et~al\mbox{.}(2020)]%
        {brown2020language}
\bibfield{author}{\bibinfo{person}{Tom Brown}, \bibinfo{person}{Benjamin Mann},
  \bibinfo{person}{Nick Ryder}, \bibinfo{person}{Melanie Subbiah},
  \bibinfo{person}{Jared~D Kaplan}, \bibinfo{person}{Prafulla Dhariwal},
  \bibinfo{person}{Arvind Neelakantan}, \bibinfo{person}{Pranav Shyam},
  \bibinfo{person}{Girish Sastry}, \bibinfo{person}{Amanda Askell},
  {et~al\mbox{.}}} \bibinfo{year}{2020}\natexlab{}.
\newblock \showarticletitle{Language models are few-shot learners}.
\newblock \bibinfo{journal}{\emph{Advances in neural information processing
  systems}}  \bibinfo{volume}{33} (\bibinfo{year}{2020}),
  \bibinfo{pages}{1877--1901}.
\newblock


\bibitem[Choi et~al\mbox{.}(2024)]%
        {choi2024malade}
\bibfield{author}{\bibinfo{person}{Jihye Choi}, \bibinfo{person}{Nils Palumbo},
  \bibinfo{person}{Prasad Chalasani}, \bibinfo{person}{Matthew~M. Engelhard},
  \bibinfo{person}{Somesh Jha}, \bibinfo{person}{Anivarya Kumar}, {and}
  \bibinfo{person}{David Page}.} \bibinfo{year}{2024}\natexlab{}.
\newblock \showarticletitle{{MALADE}: Orchestration of {LLM}-powered Agents
  with Retrieval Augmented Generation for Pharmacovigilance}. In
  \bibinfo{booktitle}{\emph{Machine Learning for Healthcare 2024}}.
\newblock


\bibitem[Cila(2022)]%
        {cila2022designing}
\bibfield{author}{\bibinfo{person}{Nazli Cila}.}
  \bibinfo{year}{2022}\natexlab{}.
\newblock \showarticletitle{Designing human-agent collaborations: Commitment,
  responsiveness, and support}. In \bibinfo{booktitle}{\emph{Proceedings of the
  2022 CHI Conference on Human Factors in Computing Systems}}.
  \bibinfo{pages}{1--18}.
\newblock


\bibitem[Clarke and Braun(2014)]%
        {clarke2014thematic}
\bibfield{author}{\bibinfo{person}{Victoria Clarke} {and}
  \bibinfo{person}{Virginia Braun}.} \bibinfo{year}{2014}\natexlab{}.
\newblock \showarticletitle{Thematic analysis}.
\newblock In \bibinfo{booktitle}{\emph{Encyclopedia of critical psychology}}.
  \bibinfo{publisher}{Springer}, \bibinfo{pages}{1947--1952}.
\newblock


\bibitem[Devlin et~al\mbox{.}(2019)]%
        {devlin2018bert}
\bibfield{author}{\bibinfo{person}{Jeff Devlin}, \bibinfo{person}{Ming-Wei
  Chang}, \bibinfo{person}{Kenton Lee}, {and} \bibinfo{person}{Kristina
  Toutanova}.} \bibinfo{year}{2019}\natexlab{}.
\newblock \showarticletitle{BERT: Pre-training of Deep Bidirectional
  Transformers for Language Understanding}. In
  \bibinfo{booktitle}{\emph{Proceedings of the 2019 Conference of the North
  {A}merican Chapter of the Association for Computational Linguistics: Human
  Language Technologies}}.
\newblock


\bibitem[Du et~al\mbox{.}(2023)]%
        {du2023improving}
\bibfield{author}{\bibinfo{person}{Yilun Du}, \bibinfo{person}{Shuang Li},
  \bibinfo{person}{Antonio Torralba}, \bibinfo{person}{Joshua~B Tenenbaum},
  {and} \bibinfo{person}{Igor Mordatch}.} \bibinfo{year}{2023}\natexlab{}.
\newblock \showarticletitle{Improving factuality and reasoning in language
  models through multiagent debate}.
\newblock \bibinfo{journal}{\emph{arXiv preprint arXiv:2305.14325}}
  (\bibinfo{year}{2023}).
\newblock


\bibitem[Gao et~al\mbox{.}(2024)]%
        {gao2024large}
\bibfield{author}{\bibinfo{person}{Chen Gao}, \bibinfo{person}{Xiaochong Lan},
  \bibinfo{person}{Nian Li}, \bibinfo{person}{Yuan Yuan},
  \bibinfo{person}{Jingtao Ding}, \bibinfo{person}{Zhilun Zhou},
  \bibinfo{person}{Fengli Xu}, {and} \bibinfo{person}{Yong Li}.}
  \bibinfo{year}{2024}\natexlab{}.
\newblock \showarticletitle{Large language models empowered agent-based
  modeling and simulation: A survey and perspectives}.
\newblock \bibinfo{journal}{\emph{Humanities and Social Sciences
  Communications}} \bibinfo{volume}{11}, \bibinfo{number}{1}
  (\bibinfo{year}{2024}), \bibinfo{pages}{1--24}.
\newblock


\bibitem[Guo et~al\mbox{.}(2024)]%
        {guo2024large}
\bibfield{author}{\bibinfo{person}{Taicheng Guo}, \bibinfo{person}{Xiuying
  Chen}, \bibinfo{person}{Yaqi Wang}, \bibinfo{person}{Ruidi Chang},
  \bibinfo{person}{Shichao Pei}, \bibinfo{person}{Nitesh~V Chawla},
  \bibinfo{person}{Olaf Wiest}, {and} \bibinfo{person}{Xiangliang Zhang}.}
  \bibinfo{year}{2024}\natexlab{}.
\newblock \showarticletitle{Large language model based multi-agents: A survey
  of progress and challenges}.
\newblock \bibinfo{journal}{\emph{arXiv preprint arXiv:2402.01680}}
  (\bibinfo{year}{2024}).
\newblock


\bibitem[Han et~al\mbox{.}(2024)]%
        {han-etal-2024-towards}
\bibfield{author}{\bibinfo{person}{Jiuzhou Han}, \bibinfo{person}{Wray
  Buntine}, {and} \bibinfo{person}{Ehsan Shareghi}.}
  \bibinfo{year}{2024}\natexlab{}.
\newblock \showarticletitle{Towards Uncertainty-Aware Language Agent}. In
  \bibinfo{booktitle}{\emph{Findings of the Association for Computational
  Linguistics: ACL 2024}}, \bibfield{editor}{\bibinfo{person}{Lun-Wei Ku},
  \bibinfo{person}{Andre Martins}, {and} \bibinfo{person}{Vivek Srikumar}}
  (Eds.). \bibinfo{publisher}{Association for Computational Linguistics},
  \bibinfo{address}{Bangkok, Thailand}, \bibinfo{pages}{6662--6685}.
\newblock
\urldef\tempurl%
\url{https://doi.org/10.18653/v1/2024.findings-acl.398}
\showDOI{\tempurl}


\bibitem[Hong et~al\mbox{.}(2024)]%
        {hong2024metagpt}
\bibfield{author}{\bibinfo{person}{Sirui Hong}, \bibinfo{person}{Mingchen
  Zhuge}, \bibinfo{person}{Jonathan Chen}, \bibinfo{person}{Xiawu Zheng},
  \bibinfo{person}{Yuheng Cheng}, \bibinfo{person}{Jinlin Wang},
  \bibinfo{person}{Ceyao Zhang}, \bibinfo{person}{Zili Wang},
  \bibinfo{person}{Steven Ka~Shing Yau}, \bibinfo{person}{Zijuan Lin},
  \bibinfo{person}{Liyang Zhou}, \bibinfo{person}{Chenyu Ran},
  \bibinfo{person}{Lingfeng Xiao}, \bibinfo{person}{Chenglin Wu}, {and}
  \bibinfo{person}{J{\"u}rgen Schmidhuber}.} \bibinfo{year}{2024}\natexlab{}.
\newblock \showarticletitle{Meta{GPT}: Meta Programming for A Multi-Agent
  Collaborative Framework}. In \bibinfo{booktitle}{\emph{The Twelfth
  International Conference on Learning Representations}}.
\newblock
\urldef\tempurl%
\url{https://openreview.net/forum?id=VtmBAGCN7o}
\showURL{%
\tempurl}


\bibitem[Hou et~al\mbox{.}({[n.\,d.]})]%
        {houdecomposing}
\bibfield{author}{\bibinfo{person}{Bairu Hou}, \bibinfo{person}{Yujian Liu},
  \bibinfo{person}{Kaizhi Qian}, \bibinfo{person}{Jacob Andreas},
  \bibinfo{person}{Shiyu Chang}, {and} \bibinfo{person}{Yang Zhang}.}
  \bibinfo{year}{[n.\,d.]}\natexlab{}.
\newblock \showarticletitle{Decomposing Uncertainty for Large Language Models
  through Input Clarification Ensembling}. In
  \bibinfo{booktitle}{\emph{Forty-first International Conference on Machine
  Learning}}.
\newblock


\bibitem[Houlsby et~al\mbox{.}(2019)]%
        {houlsby2019parameter}
\bibfield{author}{\bibinfo{person}{Neil Houlsby}, \bibinfo{person}{Andrei
  Giurgiu}, \bibinfo{person}{Stanislaw Jastrzebski}, \bibinfo{person}{Bruna
  Morrone}, \bibinfo{person}{Quentin De~Laroussilhe}, \bibinfo{person}{Andrea
  Gesmundo}, \bibinfo{person}{Mona Attariyan}, {and} \bibinfo{person}{Sylvain
  Gelly}.} \bibinfo{year}{2019}\natexlab{}.
\newblock \showarticletitle{Parameter-efficient transfer learning for NLP}. In
  \bibinfo{booktitle}{\emph{International conference on machine learning}}.
  PMLR, \bibinfo{pages}{2790--2799}.
\newblock


\bibitem[Huang et~al\mbox{.}(2023)]%
        {lapdog}
\bibfield{author}{\bibinfo{person}{Qiushi Huang}, \bibinfo{person}{Shuai Fu},
  \bibinfo{person}{Xubo Liu}, \bibinfo{person}{Wenwu Wang},
  \bibinfo{person}{Tom Ko}, \bibinfo{person}{Yu Zhang}, {and}
  \bibinfo{person}{Lilian Tang}.} \bibinfo{year}{2023}\natexlab{}.
\newblock \showarticletitle{Learning Retrieval Augmentation for Personalized
  Dialogue Generation}. In \bibinfo{booktitle}{\emph{Proceedings of the 2023
  Conference on Empirical Methods in Natural Language Processing}}.
  \bibinfo{pages}{2523--2540}.
\newblock


\bibitem[Jiang et~al\mbox{.}(2023)]%
        {jiang2023mistral}
\bibfield{author}{\bibinfo{person}{Albert~Q Jiang}, \bibinfo{person}{Alexandre
  Sablayrolles}, \bibinfo{person}{Arthur Mensch}, \bibinfo{person}{Chris
  Bamford}, \bibinfo{person}{Devendra~Singh Chaplot}, \bibinfo{person}{Diego
  de~las Casas}, \bibinfo{person}{Florian Bressand}, \bibinfo{person}{Gianna
  Lengyel}, \bibinfo{person}{Guillaume Lample}, \bibinfo{person}{Lucile
  Saulnier}, {et~al\mbox{.}}} \bibinfo{year}{2023}\natexlab{}.
\newblock \showarticletitle{{Mistral 7B}}.
\newblock \bibinfo{journal}{\emph{arXiv preprint arXiv:2310.06825}}
  (\bibinfo{year}{2023}).
\newblock


\bibitem[Kim et~al\mbox{.}(2024)]%
        {kim2024understanding}
\bibfield{author}{\bibinfo{person}{Callie~Y Kim}, \bibinfo{person}{Christine~P
  Lee}, {and} \bibinfo{person}{Bilge Mutlu}.} \bibinfo{year}{2024}\natexlab{}.
\newblock \showarticletitle{Understanding Large-Language Model (LLM)-powered
  Human-Robot Interaction}.
\newblock \bibinfo{journal}{\emph{arXiv preprint arXiv:2401.03217}}
  (\bibinfo{year}{2024}).
\newblock


\bibitem[Kuo et~al\mbox{.}(2023)]%
        {kuo2023understanding}
\bibfield{author}{\bibinfo{person}{Tzu-Sheng Kuo}, \bibinfo{person}{Hong Shen},
  \bibinfo{person}{Jisoo Geum}, \bibinfo{person}{Nev Jones},
  \bibinfo{person}{Jason~I Hong}, \bibinfo{person}{Haiyi Zhu}, {and}
  \bibinfo{person}{Kenneth Holstein}.} \bibinfo{year}{2023}\natexlab{}.
\newblock \showarticletitle{Understanding frontline workers’ and unhoused
  individuals’ perspectives on ai used in homeless services}. In
  \bibinfo{booktitle}{\emph{Proceedings of the 2023 CHI Conference on Human
  Factors in Computing Systems}}. \bibinfo{pages}{1--17}.
\newblock


\bibitem[Kwon et~al\mbox{.}(2023)]%
        {kwon2023reward}
\bibfield{author}{\bibinfo{person}{Minae Kwon}, \bibinfo{person}{Sang~Michael
  Xie}, \bibinfo{person}{Kalesha Bullard}, {and} \bibinfo{person}{Dorsa
  Sadigh}.} \bibinfo{year}{2023}\natexlab{}.
\newblock \showarticletitle{Reward design with language models}.
\newblock \bibinfo{journal}{\emph{arXiv preprint arXiv:2303.00001}}
  (\bibinfo{year}{2023}).
\newblock


\bibitem[Lee et~al\mbox{.}(2025)]%
        {lee2025veriplan}
\bibfield{author}{\bibinfo{person}{Christine Lee}, \bibinfo{person}{David
  Porfirio}, \bibinfo{person}{Xinyu~Jessica Wang}, \bibinfo{person}{Kevin
  Zhao}, {and} \bibinfo{person}{Bilge Mutlu}.} \bibinfo{year}{2025}\natexlab{}.
\newblock \showarticletitle{VeriPlan: Integrating Formal Verification and LLMs
  into End-User Planning}.
\newblock \bibinfo{journal}{\emph{arXiv preprint arXiv:2502.17898}}
  (\bibinfo{year}{2025}).
\newblock


\bibitem[Lee et~al\mbox{.}(2024)]%
        {lee2024rex}
\bibfield{author}{\bibinfo{person}{Christine~P Lee}, \bibinfo{person}{Pragathi
  Praveena}, {and} \bibinfo{person}{Bilge Mutlu}.}
  \bibinfo{year}{2024}\natexlab{}.
\newblock \showarticletitle{Rex: Designing user-centered repair and
  explanations to address robot failures}. In
  \bibinfo{booktitle}{\emph{Proceedings of the 2024 ACM designing interactive
  systems conference}}. \bibinfo{pages}{2911--2925}.
\newblock


\bibitem[Lewis et~al\mbox{.}(2020)]%
        {lewis2020retrieval}
\bibfield{author}{\bibinfo{person}{Patrick Lewis}, \bibinfo{person}{Ethan
  Perez}, \bibinfo{person}{Aleksandra Piktus}, \bibinfo{person}{Fabio Petroni},
  \bibinfo{person}{Vladimir Karpukhin}, \bibinfo{person}{Naman Goyal},
  \bibinfo{person}{Heinrich K{\"u}ttler}, \bibinfo{person}{Mike Lewis},
  \bibinfo{person}{Wen-tau Yih}, \bibinfo{person}{Tim Rockt{\"a}schel},
  {et~al\mbox{.}}} \bibinfo{year}{2020}\natexlab{}.
\newblock \showarticletitle{Retrieval-augmented generation for
  knowledge-intensive nlp tasks}.
\newblock \bibinfo{journal}{\emph{Advances in Neural Information Processing
  Systems}}  \bibinfo{volume}{33} (\bibinfo{year}{2020}),
  \bibinfo{pages}{9459--9474}.
\newblock


\bibitem[Liang et~al\mbox{.}(2023)]%
        {liang2023holistic}
\bibfield{author}{\bibinfo{person}{Percy Liang}, \bibinfo{person}{Rishi
  Bommasani}, \bibinfo{person}{Tony Lee}, \bibinfo{person}{Dimitris Tsipras},
  \bibinfo{person}{Dilara Soylu}, \bibinfo{person}{Michihiro Yasunaga},
  \bibinfo{person}{Yian Zhang}, \bibinfo{person}{Deepak Narayanan},
  \bibinfo{person}{Yuhuai Wu}, \bibinfo{person}{Ananya Kumar}, {et~al\mbox{.}}}
  \bibinfo{year}{2023}\natexlab{}.
\newblock \showarticletitle{Holistic Evaluation of Language Models}.
\newblock \bibinfo{journal}{\emph{Transactions on Machine Learning Research}}
  (\bibinfo{year}{2023}).
\newblock
\showISSN{2835-8856}
\urldef\tempurl%
\url{https://openreview.net/forum?id=iO4LZibEqW}
\showURL{%
\tempurl}
\newblock
\shownote{Featured Certification, Expert Certification}.


\bibitem[Ling et~al\mbox{.}(2024)]%
        {ling2024uncertainty}
\bibfield{author}{\bibinfo{person}{Chen Ling}, \bibinfo{person}{Xujiang Zhao},
  \bibinfo{person}{Xuchao Zhang}, \bibinfo{person}{Wei Cheng},
  \bibinfo{person}{Yanchi Liu}, \bibinfo{person}{Yiyou Sun},
  \bibinfo{person}{Mika Oishi}, \bibinfo{person}{Takao Osaki},
  \bibinfo{person}{Katsushi Matsuda}, \bibinfo{person}{Jie Ji},
  {et~al\mbox{.}}} \bibinfo{year}{2024}\natexlab{}.
\newblock \showarticletitle{Uncertainty Quantification for In-Context Learning
  of Large Language Models}. In \bibinfo{booktitle}{\emph{Proceedings of the
  2024 Conference of the North American Chapter of the Association for
  Computational Linguistics: Human Language Technologies (Volume 1: Long
  Papers)}}. \bibinfo{pages}{3357--3370}.
\newblock


\bibitem[Luo et~al\mbox{.}(2023)]%
        {luo2023empirical}
\bibfield{author}{\bibinfo{person}{Yun Luo}, \bibinfo{person}{Zhen Yang},
  \bibinfo{person}{Fandong Meng}, \bibinfo{person}{Yafu Li},
  \bibinfo{person}{Jie Zhou}, {and} \bibinfo{person}{Yue Zhang}.}
  \bibinfo{year}{2023}\natexlab{}.
\newblock \showarticletitle{An empirical study of catastrophic forgetting in
  large language models during continual fine-tuning}.
\newblock \bibinfo{journal}{\emph{arXiv preprint arXiv:2308.08747}}
  (\bibinfo{year}{2023}).
\newblock


\bibitem[Madaan et~al\mbox{.}(2024)]%
        {madaan2024selfcritic}
\bibfield{author}{\bibinfo{person}{Aman Madaan}, \bibinfo{person}{Niket
  Tandon}, \bibinfo{person}{Prakhar Gupta}, \bibinfo{person}{Skyler Hallinan},
  \bibinfo{person}{Luyu Gao}, \bibinfo{person}{Sarah Wiegreffe},
  \bibinfo{person}{Uri Alon}, \bibinfo{person}{Nouha Dziri},
  \bibinfo{person}{Shrimai Prabhumoye}, \bibinfo{person}{Yiming Yang},
  {et~al\mbox{.}}} \bibinfo{year}{2024}\natexlab{}.
\newblock \showarticletitle{Self-refine: Iterative refinement with
  self-feedback}.
\newblock \bibinfo{journal}{\emph{Advances in Neural Information Processing
  Systems}}  \bibinfo{volume}{36} (\bibinfo{year}{2024}).
\newblock


\bibitem[McDonald et~al\mbox{.}(2019)]%
        {McDonald19}
\bibfield{author}{\bibinfo{person}{Nora McDonald}, \bibinfo{person}{Sarita
  Schoenebeck}, {and} \bibinfo{person}{Andrea Forte}.}
  \bibinfo{year}{2019}\natexlab{}.
\newblock \showarticletitle{Reliability and Inter-rater Reliability in
  Qualitative Research: Norms and Guidelines for CSCW and HCI Practice}.
\newblock \bibinfo{journal}{\emph{Proceedings of the ACM on Human-Computer
  Interaction}}  \bibinfo{volume}{3} (\bibinfo{date}{11} \bibinfo{year}{2019}),
  \bibinfo{pages}{1--23}.
\newblock
\urldef\tempurl%
\url{https://doi.org/10.1145/3359174}
\showDOI{\tempurl}


\bibitem[Mikolov et~al\mbox{.}(2013)]%
        {mikolov2013distributed}
\bibfield{author}{\bibinfo{person}{Tomas Mikolov}, \bibinfo{person}{Ilya
  Sutskever}, \bibinfo{person}{Kai Chen}, \bibinfo{person}{Greg~S Corrado},
  {and} \bibinfo{person}{Jeff Dean}.} \bibinfo{year}{2013}\natexlab{}.
\newblock \showarticletitle{Distributed representations of words and phrases
  and their compositionality}.
\newblock \bibinfo{journal}{\emph{Advances in neural information processing
  systems}}  \bibinfo{volume}{26} (\bibinfo{year}{2013}).
\newblock


\bibitem[Mutlu and Forlizzi(2008)]%
        {mutlu2008robots}
\bibfield{author}{\bibinfo{person}{Bilge Mutlu} {and} \bibinfo{person}{Jodi
  Forlizzi}.} \bibinfo{year}{2008}\natexlab{}.
\newblock \showarticletitle{Robots in organizations: the role of workflow,
  social, and environmental factors in human-robot interaction}. In
  \bibinfo{booktitle}{\emph{Proceedings of the 3rd ACM/IEEE international
  conference on Human robot interaction}}. \bibinfo{pages}{287--294}.
\newblock


\bibitem[{Otter.ai}(2023)]%
        {OtterAI2023}
\bibfield{author}{\bibinfo{person}{{Otter.ai}}.}
  \bibinfo{year}{2023}\natexlab{}.
\newblock \bibinfo{title}{Otter.ai: Voice Meeting Notes}.
\newblock \bibinfo{howpublished}{\url{https://otter.ai}}.
\newblock
\newblock
\shownote{"Accessed = 03-01-2024"}.


\bibitem[Pennington et~al\mbox{.}(2014)]%
        {pennington2014glove}
\bibfield{author}{\bibinfo{person}{Jeffrey Pennington},
  \bibinfo{person}{Richard Socher}, {and} \bibinfo{person}{Christopher
  Manning}.} \bibinfo{year}{2014}\natexlab{}.
\newblock \showarticletitle{Glove: Global vectors for word representation}. In
  \bibinfo{booktitle}{\emph{Proceedings of the 2014 conference on empirical
  methods in natural language processing (EMNLP)}}.
  \bibinfo{pages}{1532--1543}.
\newblock


\bibitem[Rafailov et~al\mbox{.}(2024)]%
        {rafailov2024dpo}
\bibfield{author}{\bibinfo{person}{Rafael Rafailov}, \bibinfo{person}{Archit
  Sharma}, \bibinfo{person}{Eric Mitchell}, \bibinfo{person}{Christopher~D
  Manning}, \bibinfo{person}{Stefano Ermon}, {and} \bibinfo{person}{Chelsea
  Finn}.} \bibinfo{year}{2024}\natexlab{}.
\newblock \showarticletitle{Direct preference optimization: Your language model
  is secretly a reward model}.
\newblock \bibinfo{journal}{\emph{Advances in Neural Information Processing
  Systems}}  \bibinfo{volume}{36} (\bibinfo{year}{2024}).
\newblock


\bibitem[Ross(2001)]%
        {ross2001action}
\bibfield{author}{\bibinfo{person}{Marc~Howard Ross}.}
  \bibinfo{year}{2001}\natexlab{}.
\newblock \showarticletitle{Action evaluation in the theory and practice of
  conflict resolution}.
\newblock \bibinfo{journal}{\emph{Peace and Conflict Studies}}
  \bibinfo{volume}{8}, \bibinfo{number}{1} (\bibinfo{year}{2001}),
  \bibinfo{pages}{1--15}.
\newblock


\bibitem[Rothman(1997)]%
        {rothman1997action}
\bibfield{author}{\bibinfo{person}{Rothman}.} \bibinfo{year}{1997}\natexlab{}.
\newblock \showarticletitle{Action evaluation and conflict resolution training:
  Theory, method and case study}.
\newblock \bibinfo{journal}{\emph{International Negotiation}}
  \bibinfo{volume}{2}, \bibinfo{number}{3} (\bibinfo{year}{1997}),
  \bibinfo{pages}{451--470}.
\newblock


\bibitem[Ruan et~al\mbox{.}(2023)]%
        {ruan2023tptu}
\bibfield{author}{\bibinfo{person}{Jingqing Ruan}, \bibinfo{person}{Yihong
  Chen}, \bibinfo{person}{Bin Zhang}, \bibinfo{person}{Zhiwei Xu},
  \bibinfo{person}{Tianpeng Bao}, \bibinfo{person}{Hangyu Mao},
  \bibinfo{person}{Ziyue Li}, \bibinfo{person}{Xingyu Zeng},
  \bibinfo{person}{Rui Zhao}, {et~al\mbox{.}}} \bibinfo{year}{2023}\natexlab{}.
\newblock \showarticletitle{Tptu: Task planning and tool usage of large
  language model-based ai agents}. In \bibinfo{booktitle}{\emph{NeurIPS 2023
  Foundation Models for Decision Making Workshop}}.
\newblock


\bibitem[Salemi et~al\mbox{.}(2024)]%
        {salemi2024lamp}
\bibfield{author}{\bibinfo{person}{Alireza Salemi}, \bibinfo{person}{Sheshera
  Mysore}, \bibinfo{person}{Michael Bendersky}, {and} \bibinfo{person}{Hamed
  Zamani}.} \bibinfo{year}{2024}\natexlab{}.
\newblock \showarticletitle{LaMP: When Large Language Models Meet
  Personalization}. In \bibinfo{booktitle}{\emph{Proceedings of the 62nd Annual
  Meeting of the Association for Computational Linguistics (Volume 1: Long
  Papers)}}. \bibinfo{pages}{7370--7392}.
\newblock


\bibitem[Seeber et~al\mbox{.}(2020)]%
        {seeber2020machines}
\bibfield{author}{\bibinfo{person}{Isabella Seeber}, \bibinfo{person}{Eva
  Bittner}, \bibinfo{person}{Robert~O Briggs}, \bibinfo{person}{Triparna
  De~Vreede}, \bibinfo{person}{Gert-Jan De~Vreede}, \bibinfo{person}{Aaron
  Elkins}, \bibinfo{person}{Ronald Maier}, \bibinfo{person}{Alexander~B Merz},
  \bibinfo{person}{Sarah Oeste-Rei{\ss}}, \bibinfo{person}{Nils Randrup},
  {et~al\mbox{.}}} \bibinfo{year}{2020}\natexlab{}.
\newblock \showarticletitle{Machines as teammates: A research agenda on AI in
  team collaboration}.
\newblock \bibinfo{journal}{\emph{Information \& management}}
  \bibinfo{volume}{57}, \bibinfo{number}{2} (\bibinfo{year}{2020}),
  \bibinfo{pages}{103174}.
\newblock


\bibitem[Sullivan et~al\mbox{.}(2024)]%
        {sullivan2024making}
\bibfield{author}{\bibinfo{person}{Dakota Sullivan},
  \bibinfo{person}{Nathan~Thomas White}, \bibinfo{person}{Andrew Schoen}, {and}
  \bibinfo{person}{Bilge Mutlu}.} \bibinfo{year}{2024}\natexlab{}.
\newblock \showarticletitle{Making Informed Decisions: Supporting Cobot
  Integration Considering Business and Worker Preferences}.
\newblock \bibinfo{journal}{\emph{arXiv preprint arXiv:2401.05587}}
  (\bibinfo{year}{2024}).
\newblock


\bibitem[Thyagaraju et~al\mbox{.}(2008)]%
        {thyagaraju2008conflict}
\bibfield{author}{\bibinfo{person}{GS Thyagaraju}, \bibinfo{person}{SM Joshi},
  \bibinfo{person}{Umakanth~P Kulkarni}, \bibinfo{person}{SK NarasimhaMurthy},
  {and} \bibinfo{person}{Anil~R Yardi}.} \bibinfo{year}{2008}\natexlab{}.
\newblock \showarticletitle{Conflict resolution in multiuser context-aware
  environments}. In \bibinfo{booktitle}{\emph{2008 International Conference on
  Computational Intelligence for Modelling Control \& Automation}}. IEEE,
  \bibinfo{pages}{332--338}.
\newblock


\bibitem[Touvron et~al\mbox{.}(2023)]%
        {touvron2023llama}
\bibfield{author}{\bibinfo{person}{Hugo Touvron}, \bibinfo{person}{Thibaut
  Lavril}, \bibinfo{person}{Gautier Izacard}, \bibinfo{person}{Xavier
  Martinet}, \bibinfo{person}{Marie-Anne Lachaux},
  \bibinfo{person}{Timoth{\'e}e Lacroix}, \bibinfo{person}{Baptiste
  Rozi{\`e}re}, \bibinfo{person}{Naman Goyal}, \bibinfo{person}{Eric Hambro},
  \bibinfo{person}{Faisal Azhar}, {et~al\mbox{.}}}
  \bibinfo{year}{2023}\natexlab{}.
\newblock \showarticletitle{Llama: Open and efficient foundation language
  models}.
\newblock \bibinfo{journal}{\emph{arXiv preprint arXiv:2302.13971}}
  (\bibinfo{year}{2023}).
\newblock


\bibitem[Wu et~al\mbox{.}(2023)]%
        {wu2023autogen}
\bibfield{author}{\bibinfo{person}{Qingyun Wu}, \bibinfo{person}{Gagan Bansal},
  \bibinfo{person}{Jieyu Zhang}, \bibinfo{person}{Yiran Wu},
  \bibinfo{person}{Shaokun Zhang}, \bibinfo{person}{Erkang Zhu},
  \bibinfo{person}{Beibin Li}, \bibinfo{person}{Li Jiang},
  \bibinfo{person}{Xiaoyun Zhang}, {and} \bibinfo{person}{Chi Wang}.}
  \bibinfo{year}{2023}\natexlab{}.
\newblock \showarticletitle{Autogen: Enabling next-gen llm applications via
  multi-agent conversation framework}.
\newblock \bibinfo{journal}{\emph{arXiv preprint arXiv:2308.08155}}
  (\bibinfo{year}{2023}).
\newblock


\bibitem[xAI({[n.\,d.]})]%
        {grok}
\bibfield{author}{\bibinfo{person}{xAI}.} \bibinfo{year}{[n.\,d.]}\natexlab{}.
\newblock \showarticletitle{grok-1}.
\newblock  (\bibinfo{year}{[n.\,d.]}).
\newblock
\urldef\tempurl%
\url{https://github.com/xai-org/grok-1}
\showURL{%
\tempurl}


\bibitem[Xi et~al\mbox{.}(2023)]%
        {xi2023rise}
\bibfield{author}{\bibinfo{person}{Zhiheng Xi}, \bibinfo{person}{Wenxiang
  Chen}, \bibinfo{person}{Xin Guo}, \bibinfo{person}{Wei He},
  \bibinfo{person}{Yiwen Ding}, \bibinfo{person}{Boyang Hong},
  \bibinfo{person}{Ming Zhang}, \bibinfo{person}{Junzhe Wang},
  \bibinfo{person}{Senjie Jin}, \bibinfo{person}{Enyu Zhou}, {et~al\mbox{.}}}
  \bibinfo{year}{2023}\natexlab{}.
\newblock \showarticletitle{The rise and potential of large language model
  based agents: A survey}.
\newblock \bibinfo{journal}{\emph{arXiv preprint arXiv:2309.07864}}
  (\bibinfo{year}{2023}).
\newblock


\bibitem[Xiao et~al\mbox{.}(2023)]%
        {bge_embedding}
\bibfield{author}{\bibinfo{person}{Shitao Xiao}, \bibinfo{person}{Zheng Liu},
  \bibinfo{person}{Peitian Zhang}, {and} \bibinfo{person}{Niklas Muennighoff}.}
  \bibinfo{year}{2023}\natexlab{}.
\newblock \bibinfo{title}{C-Pack: Packaged Resources To Advance General Chinese
  Embedding}.
\newblock
\newblock
\showeprint[arxiv]{2309.07597}~[cs.CL]


\bibitem[Xie et~al\mbox{.}(2021)]%
        {xie2021explanation}
\bibfield{author}{\bibinfo{person}{Sang~Michael Xie}, \bibinfo{person}{Aditi
  Raghunathan}, \bibinfo{person}{Percy Liang}, {and} \bibinfo{person}{Tengyu
  Ma}.} \bibinfo{year}{2021}\natexlab{}.
\newblock \showarticletitle{An Explanation of In-context Learning as Implicit
  Bayesian Inference}. In \bibinfo{booktitle}{\emph{International Conference on
  Learning Representations}}.
\newblock


\bibitem[Yin et~al\mbox{.}(2024)]%
        {yin-etal-2024-reasoning}
\bibfield{author}{\bibinfo{person}{Zhangyue Yin}, \bibinfo{person}{Qiushi Sun},
  \bibinfo{person}{Qipeng Guo}, \bibinfo{person}{Zhiyuan Zeng},
  \bibinfo{person}{Xiaonan Li}, \bibinfo{person}{Junqi Dai},
  \bibinfo{person}{Qinyuan Cheng}, \bibinfo{person}{Xuanjing Huang}, {and}
  \bibinfo{person}{Xipeng Qiu}.} \bibinfo{year}{2024}\natexlab{}.
\newblock \showarticletitle{Reasoning in Flux: Enhancing Large Language Models
  Reasoning through Uncertainty-aware Adaptive Guidance}. In
  \bibinfo{booktitle}{\emph{Proceedings of the 62nd Annual Meeting of the
  Association for Computational Linguistics (Volume 1: Long Papers)}},
  \bibfield{editor}{\bibinfo{person}{Lun-Wei Ku}, \bibinfo{person}{Andre
  Martins}, {and} \bibinfo{person}{Vivek Srikumar}} (Eds.).
  \bibinfo{publisher}{Association for Computational Linguistics},
  \bibinfo{address}{Bangkok, Thailand}, \bibinfo{pages}{2401--2416}.
\newblock
\urldef\tempurl%
\url{https://doi.org/10.18653/v1/2024.acl-long.131}
\showDOI{\tempurl}


\bibitem[Yu et~al\mbox{.}(2023)]%
        {yu2023language}
\bibfield{author}{\bibinfo{person}{Wenhao Yu}, \bibinfo{person}{Nimrod
  Gileadi}, \bibinfo{person}{Chuyuan Fu}, \bibinfo{person}{Sean Kirmani},
  \bibinfo{person}{Kuang-Huei Lee}, \bibinfo{person}{Montse~Gonzalez Arenas},
  \bibinfo{person}{Hao-Tien~Lewis Chiang}, \bibinfo{person}{Tom Erez},
  \bibinfo{person}{Leonard Hasenclever}, \bibinfo{person}{Jan Humplik},
  {et~al\mbox{.}}} \bibinfo{year}{2023}\natexlab{}.
\newblock \showarticletitle{Language to rewards for robotic skill synthesis}.
\newblock \bibinfo{journal}{\emph{arXiv preprint arXiv:2306.08647}}
  (\bibinfo{year}{2023}).
\newblock


\bibitem[Zamani(2024)]%
        {rag-style}
\bibfield{author}{\bibinfo{person}{Abhiman Neelakanteswara Shreyas
  Chaudhari~Hamed Zamani}.} \bibinfo{year}{2024}\natexlab{}.
\newblock \showarticletitle{RAGs to Style: Personalizing LLMs with Style
  Embeddings}. In \bibinfo{booktitle}{\emph{The 1st Workshop on Personalization
  of Generative AI Systems}}. \bibinfo{pages}{119}.
\newblock


\bibitem[Zhang et~al\mbox{.}(2020)]%
        {zhang2020dialogpt}
\bibfield{author}{\bibinfo{person}{Yizhe Zhang}, \bibinfo{person}{Siqi Sun},
  \bibinfo{person}{Michel Galley}, \bibinfo{person}{Yen-Chun Chen},
  \bibinfo{person}{Chris Brockett}, \bibinfo{person}{Xiang Gao},
  \bibinfo{person}{Jianfeng Gao}, \bibinfo{person}{Jingjing Liu}, {and}
  \bibinfo{person}{William~B Dolan}.} \bibinfo{year}{2020}\natexlab{}.
\newblock \showarticletitle{DIALOGPT: Large-Scale Generative Pre-training for
  Conversational Response Generation}. In \bibinfo{booktitle}{\emph{Proceedings
  of the 58th Annual Meeting of the Association for Computational Linguistics:
  System Demonstrations}}. \bibinfo{pages}{270--278}.
\newblock


\bibitem[Zoom(2023)]%
        {zoom2023}
\bibfield{author}{\bibinfo{person}{Zoom}.} \bibinfo{year}{2023}\natexlab{}.
\newblock \bibinfo{title}{Video Conferencing Platform}.
\newblock
\newblock
\newblock
\shownote{"Accessed = 09-29-2023"}.


\end{thebibliography}

\appendix
\onecolumn

\section{Related Work: Personalization and Adaptation in the Era of LLMs}
\label{app:related}


In the era dominated by LLMs, encompassing commercial LLMs (\eg GPT-4~\cite{gpt-4}, Gemini families~\cite{geminiteam2023gemini}, and Claude~\cite{claude}) and ``open source'' LLMs (\eg LLaMA~\cite{touvron2023llama}, Grok~\cite{grok}, and Mistral~\cite{jiang2023mistral}), we have witnessed the remarkable capabilities of these models in generating human-like text and understanding contextual cues.
Nonetheless, just having an LLM generate answers based on its pre-training has a clear limitation: although the most-capable LLMs have been trained on large swaths of the internet, they would not have seen documents generated past a specific training cutoff date, and certainly could not have seen personal/enterprise documents, documents behind paywalls, data in databases, and so on.
To overcome such limitation, there are two common approaches for incorporating domain-specific information into LLMs for their personalized usage: fine-tuning and Retrieval Augmented Generation (RAG).


\paragraph{Adaptation with Fine-tuning.} Fine-tuning involves adapting the pre-trained LLM to a specific task or user data by continuing the training process, thereby enabling the models to learn the nuances and requirements of the target domain~\cite{houlsby2019parameter, zhang2020dialogpt, salemi2024lamp}. 
However, the fine-tuning approach requires substantial computational resources and expertise, as well as full white-box access to the model's internals, which is not a realistic assumption since LLM model weights are often protected by the model provider's intellectual property laws.
Moreover, it risks catastrophic forgetting, where the model loses its ability to perform tasks it was previously good at, and the severity of forgetting increases as the model scale increases~\cite{luo2023empirical}. 
Other than adaptation to target task, Reinforcement Learning from Human Feedback (RLHF) can help align LLMs with user preferences~\cite{bai2022rlhf}.
RLHF involves the additional steps of training a Reward Model with preference data, which consists of prompts and multiple candidate outputs ranked by human evaluators. Then it uses the reward model to fine-tune the LLM through reinforcement learning 
to align the LLM's behavior with human preferences and values defined via reward modeling.
Yet, the performance of RLHF largely depends on reward modeling and the quality and quantity of human feedback, which can be challenging to obtain, and it shares the computational expense and complexity of the fine-tuning approach. This makes fine-tuning and RLHF-based approaches less accessible to individual users who may want to leverage state-of-the-art LLMs for personal use off-the-shelf.

\paragraph{Adaptation without Fine-tuning.} On the other hand, RAG provides more flexibility to adapt to changing data without retraining the entire model, by only requiring black-box access to the model (\ie via API calls).
It allows the model to leverage knowledge sources (\eg specific documents on personal information or instructions, or an external database) not present in the original training data of the deployed LLM. 
It retrieves relevant information from the external knowledge and uses it to guide the response generation by augmenting the user input prompt~\cite{lewis2020retrieval}.  
This method is not only computationally efficient and versatile, compared to fine-tuning or RLHF approaches that necessitate additional training, but is also accessible to individuals who wish to adaptively use any black-box LLMs for personal use cases.
More importantly, it is possible with RAG to verify the validity of the generated answer by including a source citation. 

There have been a few recent attempts to explore the use of RAG in the context of personalized dialogue generation, such as persona profiles and distinctive authorial nuances~\cite{rag-style, lapdog}. 
However, the focus of personalization in our work is more complicated; it rather lies in handling conflicting multiple user requests and supporting decision-making adapted to individual preferences and instructions.

\paragraph{Beyond an LLM to LLM-powered agent.}
Other than RAG, another emerging paradigm in LLM applications is \textit{tool-use}, where LLM is instructed to produce structured outputs (\eg in JSON) that downstream code can interpret to perform tasks such as web searches, API queries, and computations~\cite{ruan2023tptu}.
By integrating RAG and tool-use, LLMs can function as agents -- an intelligent entity that can interact with each other by exchanging messages, collaborate with each other and provide feedback, just like a team of humans. 
By harnessing the collective intelligence of LLMs, multi-agent systems can tackle more complex tasks with greater reliability than a single LLM agent even with RAG or tool-use~\cite{hong2024metagpt, choi2024malade}.
To the best of our knowledge, our work is the first to extend multi-agent collaboration to multi-user adaptive personalization, moving beyond the monolithic LLM-based approaches commonly found in the prior works. 
Crucially, rather than retrofitting an existing system, we design a user-centered personalization workflow first and explicitly embed this workflow into the development of our multi-agent orchestration framework.

To the best of our knowledge, our work is the first to extend the potential of multi-agent collaboration into the domain of multi-user adaptive personalization, in contrast to the monolithic agent-based approaches found in most prior works on RAG-based personalization.
More importantly, we prioritize designing a usable personalization workflow from the users' perspectives, then integrate it explicitly into the development of our multi-agent orchestration framework. 

\section{Extended Descriptions of Multi-user Personalization Scenarios}
\label{app:scenarios-extended}

\paragraph{Scenario 1: Automated Scheduling System for Workplaces} A consulting firm utilizes an AI system to manage the scheduling of shared meeting spaces. Employees exhibit diverse preferences for these spaces---some prioritize a room for its professional ambiance and cooler climate, whereas others favor a different space due to its natural lighting and suitability for collaborative efforts. Their preferences extend to features such as whiteboards, ambient music, chairs arranged in a circular layout, and the provision of materials like sticky notes and flip charts.
Each employee follows a distinct schedule involving client consultations, team meetings, brainstorming sessions, and other engagements. Conflicts arise when multiple individuals concurrently request the same room, necessitating the introduction of a priority system. This system assigns the highest priority to client consultations, followed by team meetings, brainstorming sessions, and then all other activities, to streamline the allocation of meeting spaces.

\paragraph{Scenario 2: Assistive Care with Robot} An assistive care facility employs an AI-powered social robot to support the care of senior residents. Each resident possesses distinct preferences and schedules, necessitating tailored approaches to ensure their comfort and well-being. For instance, one resident prefers interactive wake-up calls and wants assistance for mobility in morning walks, communal breakfasts, and participation in various activities throughout the week. Conversely, another resident prioritizes independence and privacy, preferring minimal interaction with the robot and requesting that all deliveries be left at her door. 
Regarding schedules, each resident follows a personalized routine that includes unique wake-up times on different days, weekly appointments with doctors and physical therapists, participation in external activities such as yoga, art classes, and book clubs, as well as visits from family members. In instances where there are conflicts between schedules and preferences, the proposed resolution involves prioritizing requests based on the alphabetical order of the residents' first names.

\paragraph{Scenario 3: Smart-home Temperature Setting with Housemates} An AI-powered smart home device is tasked with regulating the temperature in communal areas of a residence shared by three housemates. Each housemate has personal temperature preferences and varying schedules for when these preferences should be applied. For instance, one housemate desires a warmer temperature in the study area during her hours of study or remote work. Another housemate insists on keeping the temperature at or below a certain threshold to minimize electricity costs. Meanwhile, a third housemate opts for cooler temperatures when hosting parties and warmer settings during his workout sessions in the home gym.
Furthermore, each housemate adheres to individual schedules that include work, rest, exercise, sleep, and social events throughout the week. In cases where there are conflicts between their schedules and temperature preferences, the housemates have established a consensus to resolve such issues through discussion among those involved.

\section{Implementation Details of~\oursend}
\label{app:system-details}


\subsection{Preliminaries: Design Primitives of LLM-based Agents}
\label{app:ours-primitives}
While today's LLMs exhibit impressive capabilities, they remain constrained by technical and practical limitations such as brittleness, non-determinism, limited context window, inference costs, and latency~\cite{liang2023holistic, xie2021explanation}. In other words, one cannot simply give high-level instructions to an LLM and expect it to accomplish a task, with these limitations becoming more pronounced as the complexity of the target task or application increases. 
Consequently, to best harness the capability of an LLM as a component of a complicated task, it is necessary to decompose the task into sub-tasks and manage multiple LLM conversations, each equipped with its own set of specifically-defined instructions, state, and data sources.

\paragraph{Agent}
A natural and convenient abstraction for managing such complexity is the notion of an agent who is instructed to be responsible for a specific aspect of the task.
An agent is essentially an ``intelligent'' entity that can respond to, and transform messages.
Typically, an agent encapsulates an LLM (\eg an interface to GPT-4 in our implementation) and may also be equipped with so-called \textit{tools} (also known as functions or plugins) and \textit{memory} such as conversation history or a vector database (described below).  
Much like a team of humans, agents interact by exchanging messages.
And importantly, an \textit{orchestration mechanism} is needed to manage the flow of messages between agents, to ensure that progress is made towards the completion of the task, and to handle the inevitable cases where an agent deviates from instructions. 
In this work we adopt this multi-agent paradigm, where agents are first-class citizens, acting as message transformers, and communicate by exchanging messages.

\paragraph{Tools}
As elucidated above, an LLM is essentially a text transformer; \ie in response to some input text (known as a \textit{prompt}), it produces a response. 
Free-form text responses are ideal when we want to generate a description, answer, or summary for human consumption, or even a question for another agent to answer. 
However, in some cases, we need the responses to trigger external \textit{actions}, such as a database query, web search, API call, or code execution. 
Then, we instruct the LLM response to be \textit{structured}, typically in JSON format, with various pre-specified fields, such as code, an SQL query, parameters of an API call, and so on. 
These structured responses have come to be known as tools, and the LLM is said to use a tool when it produces a structured response corresponding to a specific tool. 
To elicit a tool response from an LLM, it needs to be instructed on the expected tool format and the conditions under which it should use the tool.
Such tool-use capability allows LLM-based agents to leverage external tools and resources to accomplish tasks enhancing their functional capabilities, and operate more effectively in diverse and dynamic environments.

\paragraph{Memory}
In contrast to a solitary LLM, an agent maintains memory, which encompasses accumulated message history, detailing all queries and responses involving the agent, or external databases, facilitating in-context learning~\cite{brown2020language} to retain and retrieve information over prolonged periods. 
This capability enhances contextual coherence and fosters learning from interactions, crucial for navigating complex tasks efficiently.

As mentioned above, in constructing a system with multiple agents to tackle intricate tasks, an orchestration mechanism is critical to managing message flow between agents, ensuring task progression, and handling deviations from instructions. 
In our implementation, we build upon the Langroid library, which offers a simple yet versatile orchestration mechanism for agents, seamlessly managing user interaction, tool utilization, and sub-task delegation.

\subsection{Implementation Details}
\label{app:ours-details}


Our framework first asks the users to write down their adaptation requests or rules into documents and then preprocesses them into a database (\textit{data ingestion}).
Referring to the personalized data, the three core agents in \ours collaborate with each other to support the three phases in MAP framework: reflection, articulation, and feedback.
The reflection stage is facilitated by the \textit{Rule Manager} and \textit{Rule Retriever} agents; Rule Retriever serves as the lowest-level agent that directly extracts structured information from the database, while the medium-level agent, Rule Manager ensures that Rule Retriever extracts all users' relevant information.
At the highest level, the \textit{Planner} agent initiates interaction with the user and incorporates the hierarchical orchestration with the Rule Manager and Rule Retriever.
It handles the articulation and feedback stages seamlessly; it generates personalized responses with the guidance of the Rule Manager, addresses conflicts that may arise in satisfying all users' preferences, and provides explanations for its responses upon request. 
Below we describe the implementation details of the three agents, and tasks, tools, and resources assigned to each of them.
For an illustrative description of the overall process of \oursend, refer to Figure~\ref{fig:rag}.

\paragraph{Data Ingestion.} At the onset of the interaction, users are asked to furnish their preferences in documents.
These documents contain information pertinent to each user, encompassing individual requests, schedules, or rules. 
Subsequently, a pre-processing, or \textit{ingestion phase} ensues, wherein documents are sharded into reasonable-size chunks.
Each document chunk is mapped to a vector embedding via an embedding model (\eg BAAI/bge-large-en-v1.5~\cite{bge_embedding} in our implementation).
This process capitalizes on the inherent ability of vector embeddings to encapsulate the meaning of sentences or paragraphs~\cite{mikolov2013distributed, pennington2014glove, devlin2018bert}. 
Each vector embedding is then ingested into a vector database, along with a pointer to the chunk contents as metadata.
This database is a form of external storage where all users' personal information and preferences are stored, which is desired to be incorporated into the interactions of human users with the AI system.

\paragraph{Rule Manager}
The Rule Manager is tasked with extracting structured information from documents and presenting it in a specific nested format.
The JSON structure is explicitly defined for all users, where the highest-level fields represent users (\eg Blaine, Ryan, and Susie in Figure~\ref{fig:rag}).
Under each user field, the Rule Manager defines subfields to maximize information retrieval from the document.
Rule Manager generates questions corresponding to each field in the nested format and passes them, one by one, to the Rule Retriever who has direct access to documents (more specifically, the vector database constructed from the ingestion phase).
If the Rule Retriever cannot provide an answer to a question, the Rule Manager rephrases the question and retries a few more times until an answer is obtained. 
Once all fields in the structured format are filled, the Rule Manager returns the retrieved information to the Planner using a tool message.

\paragraph{Rule Retriever}
The Rule Retriever has direct access to documents and vector database, and is responsible for answering questions generated by the Rule Manager.
Given a question, this agent retrieves the top-$k$ most \textit{relevant} document-chunks from the vector database.
It employs a combination of two notions of relevance: (a) \textit{lexical relevance}, so-called keyword search, which is based on word-overlap between the query and the document-chunk, and (b) \textit{semantic relevance}, which is a more flexible notion of relevance based on similarity of meaning between the query and the document-chunk. 
To determine semantic relevance, the question is mapped to a vector embedding using the same embedding model as in the ingestion phase, and the top-$k$ nearest-neighbors of this vector (based on cosine similarity) are found from the vector database, and their corresponding document-chunks are retrieved. 
The original question from Rule Manager is augmented with the retrieved document-chunks, and given the augmented question, Rule Retriever provides an answer to Rule Manager.

\section{Prompt Used for Planner} \label{app:PlannerPrompt}

Here is the excerpt from the Planner's system prompt for each scenario.

\paragraph{Scenario 1: Workplace Scheduling.}
\noindent
\begin{Verbatim}[commandchars=\\\{\}, fontsize=\small]
    You are a helpful automated scheduler in my office. 
    You arrange workers’ meetings to rooms, and tell the receptionist if anything else needs to be prepared. 
    In a meeting room, workers often meet for updates, brainstorming sessions, and even little workplace birthday parties. 
    Additionally, the room is used to consult clients. 
    You cater to the individual needs and preferences of each worker, 
    ensuring a more productive and comfortable meeting environment for everyone involved.

    I will give you delivery, and you should give me (the User) personalized answers throughout the conversation.
    You MUST retrieve BOTH schedules and preferences of EVERY users, and 
    MAKE SURE that NONE of your answers conflicts with ANY of the rules and preferences that you figured out 
    based on the document.

    Workers have different preferences for setting up the room depending on who they are meeting with and 
    what they find comfortable. 
    Workers prefer the Sun room, as this is the biggest meeting room, 
    but if that is booked for a prior meeting, assign workers to use the Apple room for meetings. 
    IF CONFLICT OCCURS between workers needing the meeting rooms, PRIORITIZE giving the Sun room then the Apple room 
    in the order of client consultation, team meetings, brainstorming sessions, and then everything else. 
    If the Sun and Apple rooms are unavailable, suggest another time. 

    Once you present your answer, the I may ask for the summary of relevant rules that you referred to. 
    Then you should provide the summary of relevant rules.
\end{Verbatim}

\paragraph{Scenario 2: Assistive Care Robot}
\noindent
\begin{Verbatim}[commandchars=\\\{\}, fontsize=\small]
    You are a helpful assistive caretaking robot.
    
    I will give you delivery, and you should give me (the User) personalized answers throughout the conversation.
    You MUST retrieve ALL schedules and preferences of EVERY users, and 
    MAKE SURE that NONE of your answers conflicts with ANY of the rules and preferences that you figured out 
    based on the document.
    
    When assisting residents, if ANY resident has overlapping conflicts from their preferences 
    and schedules with another resident, serve them in ALPHABETICAL ORDER of the resident’s name. 
    Once you present your answer, the I may ask for the summary of relevant rules that you referred to. 
    Then you should provide the summary of relevant rules.
\end{Verbatim}

\paragraph{Scenario 3: Smart-home Temperature Control}
\noindent
\begin{Verbatim}[commandchars=\\\{\}, fontsize=\small]
    You are a helpful smart home agent to set temperatures in specific spaces automatically.
    Roommates all have different schedules and preferences for temperature settings. 
    Roommates have all agreed that 70 degrees would be the normal temperature setting for the house. 

    I will give you delivery, and you should give me (the User) personalized answers throughout the conversation.
    You MUST retrieve BOTH schedules and preferences of EVERY users, and 
    MAKE SURE that NONE of your answers conflicts with ANY of the rules and preferences that you figured out 
    based on the document.
    
    Once you present your answer, the I may ask for the summary of relevant rules that you referred to. 
    Then you should provide the summary of relevant rules.
\end{Verbatim}

\end{document}